\documentclass{amsart}

\theoremstyle{definition}

\theoremstyle{remark}

\numberwithin{equation}{section}

\newcommand{\bpsi}{\mbox{\boldmath $\Psi$}}
\newcommand{\q}{\mathfrak q}

\begin{document}
\title{A CLASS OF VECTOR COHERENT STATES DEFINED \\
      OVER MATRIX DOMAINS}
\author{T.Kengatharam}
\address{Department of Mathematics and Statistics, Concordia University,
   7141 Sherbrooke Street West, Montreal, Quebec H4B 1R6, Canada }
\email{t$\_$kengat@alcor.concordia.ca}
\author{S.Twareque Ali}
\address{Department of Mathematics and Statistics, Concordia University,
   7141 Sherbrooke Street West, Montreal, Quebec H4B 1R6, Canada }
\email{stali@mathstat.concordia.ca}
\date{\today}
\keywords{coherent states, vector coherent states, quaternions}
\begin{abstract}
  A general scheme is proposed for constructing vector coherent states, in analogy with  the
well-known canonical coherent states, and their deformed versions, when these latter are expressed as
infinite series in powers of a complex variable $z$. In the present scheme, the variable $z$ is replaced
by a matrix valued function over appropriate domains. As particular examples, we analyze the quaternionic
extensions of the the canonical coherent states and the Gilmore-Perelomov and Barut-Girardello coherent
states arising from representations of $SU(1,1)$.
\end{abstract}

\maketitle

\section{Introduction}

   One way to define conventional coherent states, over complex domains, is by constructing linear superpositions
$\mid z\rangle$, parametrized by a single complex number $z$, of
vectors $\{\phi_m\}_{m=0}^\infty$, which form an orthonormal basis in an infinite dimensional, complex,
separable Hilbert space $\mathfrak H$:
\begin{equation}
\mid z\rangle=\mathcal N(|z|)^{-\frac{1}{2}}\sum_{m=0}^{\infty}\frac{z^{m}}{\sqrt{\rho(m)}}\phi_{m}
\label{1}
\end{equation}
Here  $\{\rho (m)\}_{m=0}^\infty$ is a sequence of non-zero, positive numbers, chosen so as to ensure the
convergence of the sum in a non-empty open subset $\mathcal D$, of the complex plane, and $\mathcal N (|z|)$ is
a normalization constant, ensuring the
condition $\langle z \mid z\rangle = 1$. The coherent states $\mid z \rangle$ are also required to
satisfy a resolution of the identity condition:
\begin{equation}
\int_{\mathcal D}\mid z\rangle \langle z\mid d\mu = I\; ,
\label{2}
\end{equation}
where $d\mu$ is an appropriately chosen measure and $I$ the identity operator on the Hilbert space $\mathfrak H$.
These coherent states are known to have a large number of interesting properties, linking them to physical
applications, orthogonal polynomials, generalized oscillator algebras, etc. \cite{key2}, \cite{key9},
\cite{key19}.

In this paper we extend this definition to matrix domains,
thereby generating families of vector coherent states. Vector coherent states are well-known mathematical
objects, particularly when they are defined as orbits of vectors under the operators of unitary representations
of groups (see, for example, \cite{key2}, \cite{key7}, \cite{key8}, \cite{key33}). However, in the present paper
we take a completely different route for constructing them, although in special cases the link to a
group representation will also emerge.

\section{Vector coherent states -- the general set up}

  Let $\mathcal R$ be a measure space, equipped with a measure $d\mathcal R$ and $\mathcal K$ a second  measure space,
equipped with a probability measure $d\mathcal K$.
For $(r, k, \zeta ) \in \mathcal R \times \mathcal K \times [0, 2\pi)$,  let
\begin{equation}
\mathcal Z = A(r)e^{i\zeta\Theta (k)}\; ,
\label{30}
\end{equation}
where $A(r), \Theta(k)$ are two (measurable) $n\times n$ matrix-valued functions  with the following
properties (assumed to hold for almost all $r\in \mathcal R$, with respect to the measure $d\mathcal R$ and
almost all $k \in \mathcal K$, with respect to the measure $d\mathcal K$):
\begin{align}
&\Theta(k) \text{ is hermitian, that is,}\;  \Theta (k) = \Theta (k)^{\dagger}\; ,
              \label{31}\\
&\Theta(k)^{2}=\mathbb I_{n} = n\times n \;\; \text{unit matrix} \label{32}\; , \\
&[A(r),\Theta(k)]= A(r)\Theta (k) - \Theta (k) A(r) = 0\; ,  \label{33}\\
&A(r) A(r) ^{\dagger}=A(r)^{\dagger}A(r)\; .\label{34}
\end{align}
It is then straightforward to verify (e.g., by direct power series expansion) that,
\begin{equation}
\mathcal Z=A(r)e^{i\zeta\Theta(k)}=A(r)[\cos{\zeta}+i\Theta(k)\sin{\zeta}].\label{35}
\end{equation}
\label{l1}
Let $\mathcal D =  \mathcal R \times \mathcal K \times [0, 2\pi)$ and define the measure
$d\mu (r, k , \zeta ) = d\mathcal K (k)\; d\mathcal R (r)\; d\zeta$, on it.

    Let $\chi^{j}, \;\; j=1,2,...,n$, be an orthonormal basis in $\mathbb C^{n}$. Then,
$\{\chi^{j}\otimes\phi_{m}\}\;,\;\;j=1,2,...,n,\;$ $m= 0, 1,2,...,\infty \;,$ is an orthonormal basis in
$\widehat{\mathfrak H}=\mathbb C^{n}\otimes\mathfrak H$.
For each $\mathcal Z$ we define  vector coherent states (VCS) as follows:
\begin{equation}
\mid\mathcal Z,j\rangle=\mathcal N(|\mathcal Z|)^{-\frac{1}{2}}
\sum_{m=0}^{\infty}\frac{\mathcal Z^{m}}{\sqrt{\rho(m)}}\chi^{j}\otimes\phi_{m}\; , \;\; j =1,2, \ldots , n\; .
\label{36}
\end{equation}
where once again, $\mathcal N(|\mathcal Z|)$ is a normalization factor, which depends only on the
positive part $|\mathcal Z| = [\mathcal Z \mathcal Z^\dagger]^{\frac 12}$ of the matrix $\mathcal Z$,
and $\{\rho(m)\}_{m=0}^\infty$ is a sequence of
non-zero positive numbers, with $\rho (0) = 1$. These have to be chosen in a way such that the following
two conditions are satisfied.
\begin{eqnarray}
&&{\text{Normalization:}}\hspace{0.5cm}\sum_{j=0}^{n}  \langle\mathcal Z,j\mid\mathcal Z,j\rangle=1 \label{norm_cond}\\
&&{\text{Resolution of the identity:}}\hspace{0.3cm}
\sum_{j=1}^{n}\;\int_{\mathcal D}W(|\mathcal Z|)\;\mid\mathcal Z,j\rangle\langle\mathcal Z,j\mid d\mu
                  =\mathbb I_{n}\otimes I\; ,
\label{resol}
\end{eqnarray}
where $W(|\mathcal Z|)$ is an appropriately chosen positive weight function.

   A straightforward computation, using the fact that
$$\mathcal Z^{m}=A(r)^{m}e^{im\zeta\Theta(k)}=A(r)^{m}(\cos{m\zeta}+i\Theta(k)\sin{m\zeta}),$$
shows that the normalization condition (\ref{norm_cond}) implies the finiteness of the sum:
\begin{equation}
\mathcal N (\vert\mathcal Z\vert) =\sum_{m=0}^{\infty}\frac{\text{Tr}|A(r)|^{2m}}{\rho(m)}\; ,
\label{37}
\end{equation}
$|A(r)| = [A(r) A(r)^\dagger]^{\frac 12}$ denoting the positive part of the matrix $A(r)$.

    The resolution of the identity condition (\ref{resol}) imposes the following restriction on the
weight function $W(|\mathcal Z|)$ and the matrices $A(r)$:
\begin{equation}
\int_{\mathcal R}\frac{2\pi W(|\mathcal Z|)\mid A(r)\mid^{2m}}{\mathcal N(|\mathcal Z|)}\; d\mathcal R
     =\rho(m)  \mathbb I_{n}\; ,
\label{41}
\end{equation}
which can be interpreted as a sort of ``matrix moment condition''.

To see this  we note that,
\begin{align*}
&\int_{\mathcal D}W(|\mathcal Z|)\sum_{j=1}^{n}\mid\mathcal Z,j\rangle
        \langle\mathcal Z,j\mid d\mu\\
&=\int_{\mathcal D}W(|\mathcal Z|)\sum_{j=1}^{n}\;\mathcal N(\vert\mathcal Z\vert)^{-1}\;\mid
\sum_{m=0}^{\infty}\frac{\mathcal Z^{m}}{\sqrt{\rho(m)}}\chi^{j}\otimes\phi_{m}\rangle
\langle\sum_{l=0}^{\infty}\frac{\mathcal Z^{l}}{\sqrt{\rho(l)}}\chi^{j}\otimes\phi_{l}
  \mid \;d\mu\\
&=\sum_{m=0}^{\infty}\sum_{l=0}^{\infty}\int_{\mathcal D}\frac{W(|\mathcal Z|)}{\mathcal N(|\mathcal Z|)
\sqrt{\rho(m)\rho(l)}}A(r)^{m}e^{im\zeta\Theta(k)}\left(\sum_{j=1}^{n}\mid\chi^{j}\rangle\langle\chi^{j}\mid \right)\\
&\qquad \times A(r)^{l\dagger}e^{-il\zeta\Theta(k)^{\dagger}}\otimes\mid\phi_{m}\rangle
\langle\phi_{l}\mid\; d\mu\; ,\\
\end{align*}
Using
\begin{align*}
& \sum_{j=1}^{n}\mid\chi^{j}\rangle\langle\chi^{j}\mid=\mathbb I_{n},
 \qquad \Theta(k)^{\dagger}=\Theta(k)\quad {\text{and}}\\
& \int_{0}^{2\pi}e^{i(m-\ell)\zeta\Theta(k)}d\zeta=\left\{
\begin{array}{ccc}
0 & \text{if} & \ell \neq m \\
2\pi\mathbb I_n & \text{if} & \ell = m\; ,
\end{array}
\right.
\end{align*}
we reduce the last line to
\begin{align*}
&\sum_{m=0}^{\infty}\int_{\mathcal R}\int_{\mathcal K}\frac{2\pi W(|\mathcal Z|)}{\mathcal N(|\mathcal Z|)
\rho(m)}A(r)^{m}A(r)^{m\dagger}\otimes\mid\phi_{m}\rangle\langle\phi_{m}\mid d\mathcal R d\mathcal K\\
&\quad\qquad=\sum_{m=0}^{\infty}\int_{\mathcal R}\int_{\mathcal K}\frac{2\pi W(|\mathcal Z|)}{\mathcal N(|\mathcal Z|)
\rho(m)}\mid A(r)\mid^{2m}\otimes\mid\phi_{m}\rangle\langle\phi_{m}\mid d\mathcal R d\mathcal K\; .\\
\end{align*}
Since $d\mathcal K$ is a probability measure, using the fact that
$\sum_{m=0}^\infty \mid\phi_{m}\rangle\langle\phi_{m}\mid = I$ and imposing the condition (\ref{41}), we
immediately arrive at (\ref{resol}).

  There is an associated matrix-valued reproducing kernel, $K(\mathcal Z^{\dagger}, \mathcal Z')$, with
matrix elements,
\begin{align}
  K_{j\ell}(\mathcal Z^{\dagger}, \mathcal Z') & = \langle \mathcal Z , j  \mid \mathcal Z' , \ell\rangle
   = \sum_{m=0}^{\infty}\frac{1}{\rho(m)\sqrt{\mathcal N(\vert\mathcal Z\vert)\mathcal N(\vert\mathcal Z'\vert)}}\\
   &\times\langle
  e^{-im(\zeta^{'}\Theta(k')-\zeta\Theta(k))}A(r')^{m\dagger}A(r)^{m}\chi^{j}\mid\chi^{k}\rangle\nonumber.
\label{43}
\end{align}
In view of (\ref{resol}), this kernel satisfies the reproducing condition,
\begin{equation}
    \int_{\mathcal D} K(\mathcal Z^{\dagger}, \mathcal Z'')K(\mathcal {Z''}^{\dagger}, \mathcal Z')
                 \; d\mu (k'' , r'', \zeta'')= K(\mathcal Z^{\dagger}, \mathcal Z')\; .
\end{equation}

\section{ Generalized annihilation, creation and number operators}

   There are a number of operators, associated with the coherent states (\ref{1}), which define the so called
generalized oscillator algebras \cite{borzov01}, \cite{borzov95}, \cite{odz01}, \cite{odz98}.
Similar operators can also be constructed in the context of the
VCS (\ref{36}). In order to do that, let us first define
$x_{m}=\displaystyle{\frac{\rho(m)}{\rho(m-1)}}$, for $m = 1, 2, 3,
\ldots . $ Thus we write $\rho(m)= x_m x_{m-1} \ldots x_1 = x_{m}!$ and define $x_0 ! = 1$.
The generalized annihilation or lowering operator, defined on the Hilbert space $\mathfrak H$,
with respect to the basis $\{\phi_{m}\}_{m=0}^\infty$ is then written as
\begin{equation}
a\phi_{m}=\sqrt{x_{m}}\phi_{m-1}\hspace{.5cm}{\text{with}}\hspace{.5cm}a\phi_{0}=0\; .
\label{44}
\end{equation}
In the case where $x_m = m$, we recover from this the standard annihilation operator for a harmonic
oscillator. It is also easy to see that this operator acts on the coherent states $\mid z\rangle$ in
the expected manner:
$$a\mid z\rangle = z\mid z\rangle\; . $$
Using $a$, we construct the creation or raising operator $a^\dagger$ and the number
operator $N' = a^\dagger a$:
\begin{equation}
a^{\dagger}\phi_{m}=\sqrt{x_{m+1}}\phi_{m+1},\hspace{0.5cm}N'\phi_{m}=x_{m}\phi_{m}\label{45}.
\end{equation}
These three operators generate a Lie algebra (under composition given
by the commutator bracket). This is the
so-called generalized oscillator algebra, which we denote by $\mathfrak A_{\text{osc}}$. In general, the
dimension of this algebra is not finite.

  On the Hilbert space, ${\mathbb C}^n\otimes\mathfrak H$, of the VCS $\mid\mathcal Z,j\rangle$,
we define the corresponding operators as,
\begin{eqnarray}
&& A=\mathbb I_{n}\otimes a\hspace{.5cm}{\text{annihilation operator}}\label{46}\\
&& A^{\dagger}=\mathbb I_{n}\otimes a^{\dagger}\hspace{.5cm}{\text{creation operator}}\label{47}\\
&&N=\mathbb I_{n}\otimes N'\hspace{.5cm}{\text{number operator}}\label{48}.
\end{eqnarray}
They act on the VCS as
\begin{eqnarray}
A \mid\mathcal Z,j\rangle &=&\mathcal Z\mid\mathcal Z,j\rangle\; ,  \label{49}\\
A^{\dagger}\mid\mathcal Z,j\rangle&=&\mathcal N(\mathcal Z)^{-\frac{1}{2}}
\sum_{m=0}^{\infty}\sqrt{\frac {x_{m+1}}{x_m !}}\;Z^{m}\chi^{j}\otimes \phi_{m+1}\; , \label{50}\\
N \mid\mathcal Z,j\rangle &=&\mathcal N(\mathcal Z)^{-\frac{1}{2}}
\sum_{m=1}^{\infty} \frac {x_m}{\sqrt{x_m !}}\;\mathcal Z^{m}\chi^j\otimes \phi_{m}\; , \label{51}
\end{eqnarray}
and generate the Lie algebra $\mathbb I_{n} \otimes \mathfrak A_{\text{osc}}$, which again is generally
not finite dimensional.

  Using the operators $a$ and $a^\dagger$, we may also define the (formally) self-adjoint operators,
\begin{equation}
\widehat q  =\frac{a + a^{\dagger}}{\sqrt{2}}\hspace{.5cm}{\text{and}}\hspace{.5cm}
\widehat p =\frac{a - a^{\dagger}}{\sqrt{2}i},
\label{55}
\end{equation}
and the related operators
\begin{equation}
 Q  =\frac{A + A^{\dagger}}{\sqrt{2}}= \mathbb I_n \otimes \widehat q \hspace{.5cm}{\text{and}}\hspace{.5cm}
 P =\frac{A - A^{\dagger}}{\sqrt{2}i}= \mathbb I_n \otimes \widehat p\; .
\label{qp-ops}
\end{equation}
We shall need these operators later, when constructing minimal uncertainty states.

   To end this section, let us note that, as a consequence of the resolution of the identity (\ref{resol}), there
is a natural isometric embedding of the Hilbert space of the VCS into a space of vector valued functions on the
domain $\mathcal D$.  Indeed, let
$\widetilde{\mathfrak H}=L^{2}(\mathcal D , d\mu)$. Then,
\begin{equation}
\mathcal W:\mathbb C^{n}\otimes\mathfrak H\longrightarrow\mathbb C^{n}\otimes\widetilde{\mathfrak H}\hspace{.5cm}
{\text{where,}} \hspace{.5cm}(\mathcal W\Psi)^{j}(\mathcal Z)=\langle\mathcal Z,j\mid\Psi\rangle\; ,\label{64}
\end{equation}
is easily seen to be an isometry.

\section{Quaternionic canonical coherent states}
   As a first example of our general construction, we build in this section VCS using the complex representation of
quaternions by $2 \times 2$ matrices. Using the basis matrices,
$$\sigma_{0}=\left(\begin{array}{cc}
1&0\\
0&1
\end{array}\right),\;\;
i\sigma_{1}=\left(\begin{array}{cc}
0&i\\
i&0
\end{array}\right),\;\;
-i\sigma_{2}=\left(\begin{array}{cc}
0&-1\\
1&0
\end{array}\right),\;\;
i\sigma_{3}=\left(\begin{array}{cc}
i&0\\
0&-i
\end{array}\right),
$$
where $\sigma_{1},\sigma_{2}$ and $\sigma_{3}$ are the usual Pauli matrices, a general quaternion is written as
$$\ q = x_{0}\sigma_{0}+i\underline{x}\cdot\underline\sigma$$
with $x_0 \in \mathbb R , \;\; \underline{x}=(x_{1},x_{2},x_{3})\in \mathbb R^3$ and
$\underline\sigma=(\sigma_{1},-\sigma_{2},\sigma_{3})$. Thus,
\begin{equation}
   \q =\left(\begin{array}{cc}
x_{0}+ix_{3}&-x_{2}+ix_{1}\\
x_{2}+ix_{1}&x_{0}-ix_{3}
\end{array}\right)\; .\label{q3}
\end{equation}
It is convenient to introduce the polar coordinates:
$$x_{0}=r\cos{\theta},\;\; x_{1}=r\sin{\theta}\sin{\phi}\cos{\psi},\;\;x_{2}=
r\sin{\theta}\sin{\phi}\sin{\psi},\;\;x_{3}=r\sin{\theta}\cos{\phi}\; ,$$
where $r\in [0,\infty),\theta,\phi\in [0,\pi] $ and $\psi\in [0,2\pi)$. In terms of these,
\begin{equation}
\q=A(r)e^{i\theta\sigma(\widehat{n})}
\label{q4}
\end{equation}
where
\begin{equation}
A(r)=r\mathbb \sigma_0\; , \hspace{.5cm}
 \sigma(\widehat{n})=\left(\begin{array}{cc}
\cos{\phi}&\sin{\phi}e^{i\psi}\\
\sin{\phi}e^{-i\psi}&-\cos{\phi}
\end{array}\right) \hspace{.5cm} \text{and}\hspace{.5cm}\sigma(\widehat{n})^2 = \sigma_0\; .
\label{q5}
\end{equation}
We denote the field of quaternions by $\mathbb H$.

     The matrices $A(r)$ and $\sigma(\widehat{n})$ satisfy the conditions (\ref{31})-(\ref{34}).
Thus, with $\{\phi_{m}\}_{m=0}^\infty$ an orthonormal basis of an abstract Hilbert space
$\mathfrak H$ and $\chi^{1},\chi^{2}$
an orthonormal basis of $\mathbb C^{2}$, we can define the  VCS,
\begin{equation}
\mid \q, j \rangle=\mathcal N(\vert \q\vert)^{-\frac{1}{2}}\sum_{m=0}^{\infty}
\frac{\q^{m}}{\sqrt{x_m !}}\chi^{j}\otimes\phi_{m}
\in \mathbb C^2\otimes\mathfrak H\; , \hspace{.5cm} j=1,2
\label{qv2}
\end{equation}
where $\mathcal N(\vert \q\vert)$ and $x_m !$ have to be chosen appropriately.

   In order to determine the normalization constant $\mathcal N(\vert \q\vert)$, and the resolution of the identity,
first note that in order for the norm of the vector $\mid \q, j \rangle$ to be finite, we must have,
$$  \langle \q, j\mid \q, j \rangle = \mathcal N(\vert \q\vert)^{-1}
             \sum_{m=0}^\infty \frac {r^{2m}}{x_m!} < \infty\; . $$
Thus if $\lim_{m \rightarrow \infty}x_m = x$, we need to restrict $r$ to $0\leq r < L = \sqrt{x}$ for the
convergence of the above series. In this case, we define
$$ \mathcal D = \{(r, \theta , \phi , \psi )\mid 0\leq r < L\;, \;\; 0 \leq \phi \leq \pi\; , \;\;
                             0 \leq  \theta , \psi < 2\pi \}\;, $$
and note that
$$ \mathcal N (\vert \q \vert) = \mathcal N (r) = 2\sum_{m=0}^\infty \frac {r^{2m}}{x_m!}\; . $$
In the special case when $x_m =m\; , \;\; \mathcal N (\vert \q \vert)  = 2\exp[r^2]\;$,
and $\mathcal D = \mathbb R^+\times (0, 2\pi]\times S^2$, where $S^2$ is the surface of the unit two-sphere
and $(\phi , \psi)$ are the angular coordinates of a point on it.  Note that $\mathcal D$ can also be
identified with $TS^2$, the tangent bundle of $S^2$. On $\mathcal D$ we introduce the measure:
$d\mu (r, \theta , \phi , \psi )= r\;dr\;d\theta \;d\Omega (\phi , \psi )$ with
$d\Omega(\phi , \psi ) = \displaystyle{\frac 1{4\pi}}\;\sin{\phi}\;d\phi\; d\psi$.

   To obtain a resolution of the identity, we now have to find a density function $W(\vert q \vert) = W(r)$,
such that
\begin{equation}
 \int_{\mathcal D} \mid \q , j \rangle\;W(r)\;\langle \q , j \mid \; d\mu = \mathbb I_{2}\otimes I\; .
\label{qu-resolid}
\end{equation}
Since
$$\int_{0}^{2\pi}\int_{0}^{2\pi}\int_{0}^{\pi}e^{i(m-l)\theta\sigma(\widehat{n})}\sin{\phi}d\phi d\theta d\psi
=\left\{\begin{array}{ccc}
2\pi\;\mathbb I_{2}&{\text{if}}&m=l\\
0&{\text{if}}&m\not=l
\end{array}\right.$$
the moment condition (\ref{41}) becomes
$$
\int_{0}^{\infty}\frac{2\pi W(r) r^{2m+1}}{\mathcal N(r)}\; dr\; \mathbb I_{2}= x_m !\;\mathbb I_{2}\; .
$$
Writing $W(r) = \displaystyle{\frac {\mathcal N (r)}{2\pi}}\; \lambda (r)$, this is equivalent to solving the
moment problem
\begin{equation}
   \int_0^L \lambda (r) r^{2m + 1}\; dr = x_m !\; ,
\label{qu-mom-prob}
\end{equation}
for determining the auxiliary density $\lambda (r)$. With this choice of $\lambda$ the resolution of the
identity (\ref{qu-resolid}) will be satisfied. As an example, if $x_m ! = m!$ we have $L = \infty$ and then
$W(r)=\displaystyle{\frac 2\pi}$. We shall call the corresponding VCS
\begin{equation}
\mid \q, j \rangle =\frac {e^{-\frac {r^2}2}}{\sqrt{2}}\sum_{m=0}^{\infty}
\frac{q^{m}}{\sqrt{m !}}\; \chi^{j}\otimes\phi_{m}
\in \mathbb C^2\otimes\mathfrak H\;,
\label{qu-CS}
\end{equation}
{\em quaternionic canonical coherent states}. These are the natural generalizations, to quaternions,
of the well known {\em canonical coherent states} \cite{key2}.
\begin{equation}
\mid z \rangle = {e^{-\frac {r^2}2}} \sum_{m=0}^{\infty}
\frac{z^{m}}{\sqrt{m !}}\;\phi_{m}
\in \mathfrak H\;,
\label{can-CS}
\end{equation}
defined over $\mathbb C$. Treating the vectors $\mid \q, 1 \rangle$ and $\mid \q, 2 \rangle$ as elements
of a  basis, we shall define a  general quaternionic VCS as a linear combination,
\begin{equation}
\mid \q, \chi \rangle = \sum_{j=1}^2 c_j \mid \q, j \rangle, \quad \text{where} \quad c_1 , c_2 \in \mathbb C\; ,
\; \; \vert c_1\vert^2 +\vert c_2\vert^2 = 1\; , \quad \chi = \sum_{j=1}^2 c_j\chi^j\; .
\end{equation}

\section{Minimum uncertainty and analyticity properties}

    It is well-known that the canonical coherent states (\ref{can-CS}) are also states of minimum
uncertainty, in the sense that for any one of these states $\mid z\rangle$,
\begin{equation}
  \langle\Delta \widehat q\rangle_z \;\langle\Delta \widehat p\rangle_z  = \frac 12\; ,
  \qquad (\text{assuming}\;\; \hbar = 1)\; ,
\label{must1}
\end{equation}
where, for any operator $A$ on $\mathfrak H$ and any vector $\phi \in \mathfrak H$,
$$
   \langle\Delta A\rangle_\phi  = \left[ \langle\phi\mid A^2\phi\rangle -
            (\langle\phi\mid A\phi\rangle)^2\right]^\frac 12\; . $$
It is possible to construct quaternionic VCS with similar properties. To see this, first note that the matrix
$\q$ can be diagonalized as,
\begin{equation}
   \q = u (\theta , \phi )\begin{pmatrix} z & 0\\ 0 & \overline{z}\end{pmatrix}
   u (\theta , \phi )^\dagger\; ,
\label{q-diag}
\end{equation}
where,
$$  u (\theta , \phi ) =
  \begin{pmatrix} ie^{i\frac {\phi}2}\cos{\frac {\theta}2} & -e^{i\frac {\phi}2}\sin{\frac {\theta}2}\\
             e^{-i\frac {\phi}2}\sin{\frac {\theta}2} & -ie^{-i\frac {\phi}2}\cos{\frac {\theta}2}\end{pmatrix}\;
\quad \text{and} \quad z = re^{i\psi}\;. $$
Let $\chi^+ (\theta , \phi)$ and $\chi^-(\theta , \phi)$ be the two (normalized) eigenvectors,
corresponding to the eigenvalues $z$ and
$\overline{z}$, respectively. Define the two quaternionic VCS,
\begin{align}
  \mid \q , + \rangle  & = e^{-\frac {r^2}2}\;\sum_{m=0}^{\infty}
\frac {\q^{m}}{\sqrt{m !}}\; \chi^{+}(\theta , \phi)\otimes\phi_{m} = e^{-\frac {r^2}2}\;\sum_{m=0}^{\infty}
\frac {z^m}{\sqrt{m !}}\; \chi^{+}(\theta , \phi)\otimes\phi_{m}\; ,\label{must2}\\
 \mid \q , - \rangle  & = e^{-\frac {r^2}2}\;\sum_{m=0}^{\infty}
\frac {\q^{m}}{\sqrt{m !}}\; \chi^{-}(\theta , \phi)\otimes\phi_{m} = e^{-\frac {r^2}2}\;\sum_{m=0}^{\infty}
\frac{\overline{z}^m}{\sqrt{m !}}\; \chi^{-}(\theta , \phi)\otimes\phi_{m}\; .\label{must3}
\end{align}
The normalization of these states has been chosen to ensure that $\langle \q, \pm\;\vert\; \q, \pm \rangle =1$.
From the nature of the operators $Q$ and $P$, defined in (\ref{qp-ops}), it is then clear that these states also
have minimum uncertainty:
\begin{equation}
  \langle\Delta Q\rangle_\pm \;\langle\Delta P\rangle_\pm  = \frac 12 \; .
\label{must4}
\end{equation}

  Next let us look a little more closely at the nature of the isometry (\ref{64}), for the quaternionic VCS. Recall
that in this case, $\mathcal D = \mathbb R^+\times (0, 2\pi]\times S^2 \simeq TS^2$. Once again, let
$\widetilde{\mathfrak H}= L^{2}(\mathcal D, d\mu)$. We are interested in the isometry
\begin{equation}
\mathcal W:\mathbb C^{2}\otimes\mathfrak H\longrightarrow \mathbb C^{2}\otimes\widetilde{\mathfrak H}
\hspace{.5cm}{\text{with}}\hspace{.5cm}(\mathcal W\bpsi)^{j}(\q)=\langle\bpsi\mid \q, j\rangle.
\label{qv222}
\end{equation}
A general vector $\bpsi\in \mathbb C^2\otimes\mathfrak H$ has the form, $\bpsi = \sum_{j=1}^2\chi^j\psi_j$, with
$\psi_j \in \mathfrak H$. We write $\mathbf F = \mathcal W\bpsi$ and introduce the functions,
\begin{align*}
   & f_j (z) =  \sum_{m=0}^\infty\frac {z^m}{\sqrt{m !}}\;\langle\phi_m\vert\psi_j\rangle_{\mathfrak H}\; , \;\;
   j=1,2\; ,
    \qquad \mathbf f (z) = \sum_{j = 1}^2\chi^j f_j (z)\; , \\
   & f_j (\overline{z})  =  \sum_{m=0}^\infty
        \frac {\overline{z}^m}{\sqrt{m !}}\;\langle\phi_m\vert\psi_j\rangle_{\mathfrak H}\; ,\;\; j=1,2\; ,
        \qquad
        \mathbf f (\overline{z}) = \sum_{j = 1}^2\chi^j f_j (\overline{z}) \; .
\end{align*}
A straightforward computation then shows that the image of the isometry (\ref{qv222}) consists of vector
valued functions of the type,
\begin{equation}
   \mathbf F (z, \overline{z}, \theta , \phi) = \frac{1}{\sqrt{2}}e^{-\frac{\vert z\vert^{2}}{2}}\left[
        \mathbb P^{+}(\theta , \phi )\mathbf f (\overline{z})  +
                      \mathbb P^{-}(\theta , \phi )\mathbf f (z)\right]\; .
\label{holfunct}
\end{equation}
where $\mathbb P^{\pm}(\theta , \phi )$ are  eigenprojectors corresponding to the eigenvectors
$\chi^\pm(\theta , \phi )$, respectively.
Thus, for fixed $(\theta , \phi)$, each component function $F^j (z, \overline{z}, \theta , \phi)$
is a linear combination
of two holomorphic functions $f_1 (z), f_2 (z)$ and their antiholomorphic counterparts.

\section{Relation to the Weyl-Heisenberg group}

   The canonical coherent states (\ref{can-CS}) can be expressed (see, for example, \cite{key2}) in the form,
\begin{equation}
   \vert z \rangle = e^{za^\dagger - \overline{z}a}\phi_0 = e^{i(p\widehat q - q\widehat p)}\phi_0 \;,
  \quad \text{where} \quad z = \frac {q-ip}{\sqrt{2}}\; .
\label{ccs-gr1}
\end{equation}
We now show that the quaternionic canonical coherent states (\ref{qu-CS}) also have the  analogous representation:
\begin{equation}
\mid \q , j\rangle=\frac{1}{\sqrt{2}}e^{\q\otimes a^{\dagger} - \q^{\dagger}\otimes a}\chi^{j}\otimes\phi_{0}
\label{qv28}
\end{equation}\label{tq1}
To see this, note that,
\begin{equation*}
[\q^\dagger\otimes a,\;\q\otimes a^\dagger ] = r^{2}\mathbb I_{2}\otimes I\; .
\end{equation*}
Next, since for  two operators $A$ and $B$, the commutator of which commutes with both $A$ and $B$, the
Baker-Campbell-Hausdorff identity,
$$e^{A+B}=e^{-\frac{1}{2}[A,B]}e^{A}e^{B}\; , $$
holds, we may write
$$e^{\q\otimes a^{\dagger} - \q^{\dagger}\otimes a}=
e^{-\frac{1}{2}[\q\otimes a^{\dagger},\; - \q^{\dagger}\otimes a]}
e^{\q\otimes a^{\dagger}}e^{- \q^{\dagger}\otimes a}.$$
Since $a^{m}\phi_{0}=0$ for all $m\geq 1$, we have
$$e^{- \q\otimes a}\chi^{j}\otimes\phi_{0}=\chi^{j}\otimes\phi_{0}$$
and
\begin{eqnarray*}
e^{\q\otimes a^{\dagger}}(\chi^{j}\otimes\phi_{0})&=&\sum_{m=0}^{\infty}\frac{(\q\otimes a^{\dagger})^{m}}{m!}
         \chi^{j}\otimes\phi_{0}
   =\sum_{m=0}^{\infty}\frac{\q^{m}\chi^{j}\otimes a^{\dagger m}\phi_{0}}{m!}\\
&=&\sum_{m=0}^{\infty}\frac{\q^{m}}{\sqrt{m!}}\chi^{j}\otimes\phi_{m}
\end{eqnarray*}
Thus,
$$\frac{1}{\sqrt{2}}e^{\q\otimes a^{\dagger} - \q^{\dagger}\otimes a}\chi^{j}\otimes\phi_{0}=
\frac{1}{\sqrt{2}}e^{-\frac{r^{2}}{2}}\sum_{m=0}^{\infty}\frac{\q^{m}}{\sqrt{m!}}\chi^{j}
\otimes\phi_{m}=\mid \q , j \rangle\; .$$

   To develop a group theoretical interpretation for the quaternionic canonical coherent states,
we go back to the canonical coherent
states as written out in (\ref{ccs-gr1}). The operators $\widehat q , \widehat p$ and $I$ generate an irreducible
representation of $\mathfrak g_{\text{W-H}}$, the Lie algebra of the Weyl-Heisenberg group $G_{\text{W-H}}$, on the
Hilbert space $\mathfrak H$. A unitary irreducible representation of $G_{\text{W-H}}$ on $\mathfrak H$ is given
by the operators $U(\vartheta , q, p) = e^{i(\vartheta I + p\widehat q - q\widehat p)}$. Thus, $\mid z \rangle
= U(0 , q, p)\phi_0$. Turning now to the quaternionic canonical coherent states,
as expressed in (\ref{qv28}), we find using
(\ref{q-diag}),
$$ \q\otimes a^\dagger - \q^\dagger \otimes a = u(\theta , \phi )
  \begin{pmatrix} z a^\dagger - \overline{z}a & 0\\ 0 & \overline{z}a^\dagger - za\end{pmatrix}
   u(\theta , \phi )^\dagger\; . $$
Thus,
\begin{equation}
  e^{\q\otimes a^\dagger - \q^\dagger \otimes a } = u(\theta , \phi )\begin{pmatrix} U(0, q, p ) & 0 \\
              0 & U(0, q, -p )\end{pmatrix}u(\theta , \phi )^\dagger.
\end{equation}
Writing
\begin{equation}
  \widetilde{U} (\vartheta , \q ) = \widetilde{U} (\vartheta , q , p , \theta , \phi ) :=
      u(\theta , \phi )\begin{pmatrix} U(\vartheta, q, p ) & 0 \\
              0 & U(\vartheta, q, - p )\end{pmatrix}u(\theta , \phi )^\dagger\;,
\label{q-wh-rep}
\end{equation}
we observe that for fixed $(\theta , \phi)$ these operators realize a unitary (reducible) representation of
$G_{\text{W-H}}$ on $\mathbb C^2 \otimes \mathfrak H$. In terms of these operators,
\begin{equation}
   \mid \q, j\rangle = \frac 1{\sqrt{2}}\;\widetilde{U} (0 , \q )\chi^j \otimes \phi_0 =
   \frac 1{\sqrt{2}}\;\widetilde{U} (0 , q , p , \theta , \phi )\chi^j \otimes \phi_0 \; ,
\label{qu-vcs4}
\end{equation}
in complete analogy with the case of the canonical coherent states.

\section{Quaternionic VCS from $SU(1,1)$ representations}

    As a second example of the construction of VCS using quaternions, we shall obtain, in this section,
analogues of the Gilmore-Perelomov \cite{key16}, \cite{key30}  and  Barut-Girardello \cite{{bar-gir71}}
coherent states. Both these families of  states arise from the discrete
series representations of $SU(1,1)$.
Writing $\mathcal D_1 = \{z \in \mathbb C\mid \vert z \vert < 1 \}$, the Gilmore-Perelomov  coherent states,
labelled by points of $\mathcal D_1$, are defined to be,
\begin{equation}
 \mid z;\; \mbox{\tiny G-P}\rangle = (1 - r^2 )^\kappa\; \sum_{m=0}^\infty
     \left[ \frac {(2\kappa)_m}{m!}\right]^\frac 12 z^m \phi_m \in
          \mathfrak H \;, \quad  r = \vert z \vert \;, \; \; \kappa = 1, \frac 32 , 2, \frac 52 , \ldots\;  ,
\label{su11-cs1}
\end{equation}
where we have used the Pochhammer symbol,
$$  (a)_m = \frac {\Gamma (a + m )}{\Gamma (a)} = a (a+1)(a+2)\dots (a+m-1)\; ,$$
and as before, the $\phi_m$ constitute an orthonormal basis of the Hilbert space $\mathfrak H$. The index $\kappa$
labels the unitary irreducible representation of $SU(1,1)$, to which the above coherent states are associated.
This representation is carried by the Hilbert space $\mathfrak H_{\text{hol}}(\mathcal D_1)$, which is the
subspace of all holomorphic functions in $L^2 (\mathcal D_1 , (2\kappa - 1)d\mu_\kappa )$, where
$$
   d\mu_\kappa (z, \overline{z} ) = \frac {(1 -r^2)^{2\kappa-2}}\pi\; r\;dr\;d\theta\; , \quad z = re^{i\theta}\; .$$
An element $g \in SU(1,1)$ is a complex $2\times 2$ matrix,
$$
 g =\begin{pmatrix} \alpha & \beta \\ \overline{\beta} & \overline{\alpha} \end{pmatrix}\; ,
   \quad \text{det}\; g = \vert\alpha\vert^2 - \vert\beta\vert^2 = 1\; , $$
 and the unitary irreducible representation $U^\kappa$, labelled by $\kappa$, acts on vectors $f \in
\mathfrak H_{\text{hol}}(\mathcal D_1)$ in the manner
$$
  (U^\kappa (g)f)(z) = (\alpha -\overline{\beta}z )^{-2\kappa}\;
     f\left(\frac {\overline{\alpha}z - \beta}{\alpha - \overline{\beta}z}\right)\; . $$
The monomials
$$ u_m (z) = \left[ \frac {(2\kappa)_m}{m!}\right]^\frac 12 z^m \; , $$
form an orthonormal basis in $\mathfrak H_{\text{hol}}(\mathcal D_1)$. Moreover, identifying the abstract
Hilbert space $\mathfrak H$ with $\mathfrak H_{\text{hol}}(\mathcal D_1)$ and $\phi_m$ with $u_m$,
it can be shown \cite{key2} that the coherent states (\ref{su11-cs1}) can also be written in the form
\begin{equation}
  \mid z;\; \mbox{\tiny G-P}\rangle = U^\kappa (\mathcal Z)\phi_0 \; , \quad \text{where,} \quad \mathcal Z =
     \frac 1{\sqrt{1-r^2}}\; \begin{pmatrix} 1 & z \\ \overline{z} & 1 \end{pmatrix} \in SU(1,1)\; .
\label{su11-cs2}
\end{equation}
Observe, that in the notation introduced in (\ref{1}), in this case we have,
$$ \mathcal N (\vert z \vert ) = (1 - r^2 )^{-2\kappa}, \quad \rho (m) = \left[ \frac {(2\kappa)_m}{m!}\right]^{-1}\;.$$
Thus, $x_m = \displaystyle{\frac m{2\kappa + m -1}}$ and since $\lim_{m \rightarrow \infty}x_m = 1$, this determines the
radius of convergence of the infinite series in (\ref{su11-cs1}) and hence the appearance of the unit
disc.

   The coherent states (\ref{su11-cs1}) satisfy the resolution of the identity,
\begin{equation}
\frac {2\kappa - 1}\pi\; \int_{\mathcal D_1} \mid z;\; \mbox{\tiny G-P} \rangle\langle z;\; \mbox{\tiny G-P}\mid \;
     \frac {r\;dr\;d\theta}{(1-r^2)^2} = I\; .
\label{su11-resolid}
\end{equation}
The representation of the Lie algebra of $SU(1,1)$ on  $\mathfrak H_{\text{hol}}(\mathcal D_1)$
is generated by the three operators $K_{+},K_{-}$ and $K_{3}$, which satisfy the commutation relations
\begin{equation}
   [K_{3},K_{\pm}]= \pm K_{\pm},\qquad [K_{-},K_{+}]=2K_{3}\; .
\label{su11-comm}
\end{equation}
They act on the vectors $\phi_m$ in the manner,
\begin{equation}
   K_{-}\phi_{m}  =  \sqrt{m(2\kappa + m -1)}\phi_{m-1}\; , \quad K_{+} = K_{-}^\dagger\; ,
\quad  K_{3}\phi_{m}  = (\kappa + m)\phi_{m}.
\label{su11-comm-op}
\end{equation}
Thus $K_{-}\phi_0 = 0$ and
$$
   \phi_m = \frac 1{\sqrt{m! (2\kappa )_m}}\;K_{+}^m \phi_0\; .$$
Furthermore, it can be shown \cite{fujii01} that,
\begin{equation}
 \mid z ;\; \mbox{\tiny G-P}\rangle = e^{wK_{+} - \overline{w}K_{-}}\phi_0\; , \quad w \in \mathbb C\; ,
\label{su11-cs5}
\end{equation}
where $z$ and $w$ are related by
\begin{equation}
z = \frac {w \tanh (\vert w\vert )}{\vert w \vert}\; .
\label{disent1}
\end{equation}
Equation (\ref{su11-cs5}) should be compared to (\ref{ccs-gr1}). Note however, that unlike in that case, the operators
$K_+$ and $K_-$, appearing in (\ref{su11-cs5}) are not the creation and annihilation operators naturally
associated with the expansion in (\ref{su11-cs1}) (see (\ref{44})). Indeed, in the present case the operator $a$
has the form:
\begin{equation}
 a_{\mbox{\tiny G-P}}\mid z ;\; \mbox{\tiny G-P}\rangle = z \mid z;\; \mbox{\tiny G-P}\rangle\; ,
\qquad a_{\mbox{\tiny G-P}}\phi_m = \sqrt{\frac m{2\kappa +m -1}}\phi_{m-1}\; .
\label{gp-ann-op}
\end{equation}
On the other hand, it is possible to define \cite{bar-gir71} a second set of coherent states
$\mid w;\; \mbox{\tiny B-G} \rangle$ for this same
representation of $SU(1,1)$, using $K_{-}$ as the generalized annihilation operator:
\begin{equation}
   K_{-}\mid w ;\; \mbox{\tiny B-G}\rangle := a_{\mbox{\tiny B-G}}\mid w ;\; \mbox{\tiny B-G}\rangle
     = w\mid w;\; \mbox{\tiny B-G}\rangle\; , \qquad w \in \mathbb C\; .
\label{bg-ann-op}
\end{equation}
These states, known as the Barut-Girardello coherent states, are defined for all $w \in \mathbb C$ and
they are of the form:
\begin{equation}
  \mid w;\; \mbox{\tiny B-G}\rangle = \frac {\vert w\vert^{2\kappa -1}}{\sqrt{I_{2\kappa -1}(2\vert w\vert )}}\;
     \sum_{m=0}^\infty\frac {w^m}{\sqrt{m! (2\kappa + m -1)!}}\; \phi_m \; ,
\label{bar-gir-cs}
\end{equation}
where $I_\nu (x)$ is the order $\nu$ modified Bessel function of the first kind. These
coherent states satisfy the resolution of the identity,
\begin{equation}
   \frac 2\pi\; \int_{\mathbb C}\mid w;\; \mbox{\tiny B-G}\rangle
   \langle w;\; \mbox{\tiny B-G}\mid\; K_{2\kappa -1}(2\varrho )\; I_{2\kappa -1}
          (2\varrho )\; \varrho\; d\varrho\; d\vartheta\; , \qquad w = \varrho e^{i\vartheta}\; ,
\label{bar-gir-resolid}
\end{equation}
where again, $K_\nu (x)$ is the order $\nu$ modified Bessel function of the second kind.

     It is now straightforward to write down quaternionic VCS which extend (\ref{su11-cs1}):
\begin{equation}
\mid \q, j;\; \mbox{\tiny G-P}  \rangle = \frac {(1 - r^2 )^\kappa}{\sqrt{2}}\; \sum_{m=0}^\infty
       \left[ \frac {(2\kappa)_m}{m!}\right]^\frac 12 \q^m \chi^j \otimes \phi_m \;,
          \quad  r = \vert \q \vert = [\q\q^\dagger]^\frac 12\; ,
\label{su11-vcs1}
\end{equation}
where $\q$ is a quaternionic variable with domain $\mathcal D_1 \times S^2$, and a
similar set of VCS  extending (\ref{bar-gir-cs}):
\begin{equation}
  \mid \mathfrak w, j ;\; \mbox{\tiny B-G}\rangle = \frac {r^{2\kappa -1}}{\sqrt{2\;I_{2\kappa -1}(2r)}}\;
     \sum_{m=0}^\infty\frac {\mathfrak w^m}{\sqrt{m! (2\kappa + m -1)!}}\;\chi^j\otimes \phi_m \; , \quad
       r = \vert \mathfrak w \vert \; ,
\label{bar-gir-vcs}
\end{equation}
the quaternionic variable $\mathfrak w$  being defined over the domain $TS^2$.

   In the case of the vectors (\ref{su11-vcs1}), it is also possible, using (\ref{su11-cs2}), to give a representation
theoretic interpretation along the lines of (\ref{qv28}) -- (\ref{qu-vcs4}). Indeed, by virtue of (\ref{su11-cs1}),
(\ref{su11-cs2}) and the decomposition (\ref{q-diag}) of the quaternion $\q$, we can immediately rewrite
(\ref{su11-vcs1}) as
$$
  \mid \q, j;\; \mbox{\tiny G-P}\rangle = \frac 1{\sqrt{2}}\;u(\theta , \phi )
    \begin{pmatrix} U^\kappa (\mathcal Z ) & 0 \\
     0 & U^\kappa (\mathcal Z^\dagger)\end{pmatrix}u(\theta , \phi )^\dagger\;\chi^j\otimes\phi_0 .$$
Writing
$$ \widetilde{U}^\kappa (\q) = u(\theta , \phi )\begin{pmatrix} U^\kappa (\mathcal Z ) & 0 \\
     0 & U^\kappa (\mathcal Z^\dagger)\end{pmatrix}u(\theta , \phi )^\dagger\;, $$
this immediately yields,
\begin{equation}
  \mid \q, j ;\; \mbox{\tiny G-P}\rangle = \frac 1{\sqrt{2}}\;\widetilde{U}^\kappa (\q)\chi^j \otimes \phi_0\; ,
\label{su11-vcs7}
\end{equation}
which is the analogue of (\ref{qu-vcs4}). Moreover, since by (\ref{su11-cs5}),
$$
  \widetilde{U}^\kappa (\q)\chi^j \otimes \phi_0 = u(\theta , \phi )\begin{pmatrix} e^{wK_+ -\overline{w}K_-} & 0 \\
     0 &  e^{\overline{w}K_+ -wK_-}\end{pmatrix}u(\theta , \phi )^\dagger\; \chi^j\otimes\phi_0\; $$
with $z$ and $w$ being related by (\ref{disent1}), we can now immediately transform this to
\begin{equation}
    \mid \q, j ;\; \mbox{\tiny G-P}\rangle = \frac 1{\sqrt{2}}\;
        e^{\mathfrak w \;\otimes K_+ -\mathfrak w^\dagger\otimes K_-}\chi^j\otimes \phi_0\; ,
\label{su11-vcs9}
\end{equation}
where now the quaternionic variables  $\q$ and $\mathfrak w$ are related by
\begin{equation}
\q = \frac {\mathfrak w \tanh (\vert \mathfrak w \vert )}{\vert \mathfrak w \vert}\; .
\label{disent2}
\end{equation}
Note that while $0 \leq \vert \q \vert < 1$, for the transformed variable $\mathfrak w$ we have,
$0 \leq \vert \mathfrak w \vert < \infty$.

    Interestingly, there is yet another family of coherent states, again related to the
$SU(1,1)$ group, which can be constructed using the two number operators,
$$
  N_{\mbox{\tiny G-P}} = a^\dagger_{\mbox{\tiny G-P}}\; a_{\mbox{\tiny G-P}} \quad \text{and} \quad
       N_{\mbox{\tiny B-G}} = a^\dagger_{\mbox{\tiny B-G}}\; a_{\mbox{\tiny B-G}}\; .
$$
Indeed, from (\ref{gp-ann-op}) and (\ref{bg-ann-op}),
\begin{equation}
    N_{\mbox{\tiny G-P}}\phi_m = \frac m{2\kappa +m-1}\phi_m \quad \text{and} \quad
    N_{\mbox{\tiny B-G}}\phi_m = m(2\kappa +m-1)\phi_m\; .
\label{num-ops}
\end{equation}
Thus we define a third number operator $N_{\mbox{\tiny INT}}$, essentially as one which
interpolates between these two:
\begin{equation}
  N_{\mbox{\tiny G-P}} N_{\mbox{\tiny INT}} = N_{\mbox{\tiny B-G}} \quad \Longrightarrow
      N_{\mbox{\tiny INT}} \phi_m = (2\kappa + m -1)^2 \phi_m\; ,
\label{int-num-op}
\end{equation}
and the related annihilation operator,
\begin{equation}
    a_{\mbox{\tiny INT}}\phi_m = (2\kappa +m -1)\phi_{m-1}
\label{int-ann-op}
\end{equation}
The corresponding coherent states, defined for all $w \in \mathbb C$, are
\begin{equation}
  \mid w ;\;\mbox{\tiny INT}\rangle = \mathcal N (r)^{-\frac 12}\sum_{m=0}^\infty\frac {w^m}{(2\kappa + m -1)!}
       \phi_m\; ,
       \qquad r = \vert w \vert\; ,
\label{int-cs}
\end{equation}
where the normalization constant is given by
$$ \qquad \mathcal N (r) = \frac {_1F_2 (1;\; 2\kappa , 2\kappa ; \; r^2 )}{[\Gamma (2\kappa )]^2}\; , $$
in terms of the hypergeometric function
$$
   _1F_2 (a;\; b,c;\; x) = \sum_{m=0}^\infty \frac {(a)_m}{(b)_m\; (c)_m}\cdot \frac {x^m}{m!}\; .$$
The moment problem for determining the resolution of the identity is now
$$ \pi\int_0^\infty r^m \lambda (r) \; dr = [(2\kappa + m -1)!]^2\; .$$
This can be explicitly solved to yield,
$$
    \lambda (r) = \frac {2}\pi r^{2\kappa-1} K_0 (2\sqrt{r})\; , $$
where once again, $K_0$ is the order-0 modified Bessel function of the second kind. Finally, one obtains
\begin{equation}
  \int_{\mathbb C}\mid w ;\;\mbox{\tiny INT}\rangle\langle w ;\;\mbox{\tiny INT}\mid\; d\mu_{\mbox{\tiny INT}}
                  (w , \overline{w}) = I\;,
\label{int-resolid1}
\end{equation}
with
\begin{equation}
  d\mu_{\mbox{\tiny INT}}(w , \overline{w}) = \frac {2r^{4\kappa -1}}{\pi [\Gamma (2\kappa )]^2}\; K_0 (2r)\;
   _1F_2 (1;\; 2\kappa , 2\kappa ; \; r^2 )\; d\theta\; dr\; .
\label{int-resolid2}
\end{equation}
The corresponding quaternionic VCS are then
\begin{equation}
\mid \mathfrak w , j ;\;\mbox{\tiny INT}\rangle =
   \frac 1{\sqrt{2}}\;\frac {\Gamma (2\kappa )}{[ _1F_2 (1;\; 2\kappa , 2\kappa ; \; r^2 )]^\frac 12}
        \sum_{m=0}^\infty\frac {\mathfrak w^m}{(2\kappa + m -1)!}\chi^j\otimes
       \phi_m\; ,
\label{int-qu-vcs}
\end{equation}
where $r = \vert \mathfrak w \vert$ and   $\mathfrak w \in TS^2\; .$

   The fact that the coherent states (\ref{int-cs}) are indeed related to the $SU(1,1)$ group is brought out
more clearly by the following observation: computing the commutator
$[a_{\mbox{\tiny INT}},\; a_{\mbox{\tiny INT}}^\dagger ]$ we find,
$$
  [a_{\mbox{\tiny INT}},\; a_{\mbox{\tiny INT}}^\dagger ]\phi_m = [2(2\kappa +m ) -1]\phi_m\; $$
Let us define a new ``number operator'' $\widetilde{N}_{\mbox{\tiny INT}} $ by the action
\begin{equation}
    \widetilde{N}_{\mbox{\tiny INT}}\phi_m = (2\kappa + m - \frac 12 )\phi_m \; ,
\label{new-num-op}
\end{equation}
on the basis vectors $\phi_m$. Then we easily establish the commutation relations,
\begin{equation}
[a_{\mbox{\tiny INT}},\; a_{\mbox{\tiny INT}}^\dagger ] = 2\widetilde{N}_{\mbox{\tiny INT}}\; , \quad
[\widetilde{N}_{\mbox{\tiny INT}},\; a_{\mbox{\tiny INT}}^\dagger ] = a_{\mbox{\tiny INT}}^\dagger\; , \quad
[\widetilde{N}_{\mbox{\tiny INT}},\; a_{\mbox{\tiny INT}} ] = - a_{\mbox{\tiny INT}}\; .
\label{su11-comm2}
\end{equation}
Comparing with (\ref{su11-comm}), we find that the three operators $a_{\mbox{\tiny INT}}\; ,
a_{\mbox{\tiny INT}}^\dagger$ and $\widetilde{N}_{\mbox{\tiny INT}}$ satisfy exactly the same commutation relation
as the three generators, $K_- \; , K_+$ and $K_3$ of $\mathfrak{su}(1,1)$, the Lie algebra of $SU(1,1)$.
Thus, they also realize
a representation of this algebra on $\mathfrak H$. The two number operators $\widetilde{N}_{\mbox{\tiny INT}}$
and $N_{\mbox{\tiny INT}}$ are related as,
\begin{equation}
   N_{\mbox{\tiny INT}} = \widetilde{N}_{\mbox{\tiny INT}}^2 - \widetilde{N}_{\mbox{\tiny INT}} + \frac 14
   = [ \widetilde{N}_{\mbox{\tiny INT}} - \frac 12 ]^2\; .
\label{num-op-reln}
\end{equation}
A similar situation was seen to arise in the case of temporally stable coherent states related to
the infinite well and P\"oschl-Teller potentials, where the Lie algebra  $\mathfrak{su}(1,1)$ appeared as a dynamical
algebra. It ought to be pointed out, however, that the representation of $\mathfrak{su}(1,1)$, generated by the
operators $K_\pm , K_3$ in (\ref{su11-comm})-(\ref{su11-comm-op}), is different from the one generated by the
operators $a_{\mbox{\tiny INT}}^\dagger ,\; a_{\mbox{\tiny INT}}$ and $\widetilde{N}_{\mbox{\tiny INT}}$. Indeed,
computing the Casimir operators in the two cases, we find that $\frac 12 (K_- K_+ + K_+ K_- ) - K_3^2 =
\kappa (1 - \kappa)$ while, $\frac 12 (a_{\mbox{\tiny INT}}a_{\mbox{\tiny INT}}^\dagger +
a_{\mbox{\tiny INT}}^\dagger a_{\mbox{\tiny INT}} ) - \widetilde{N}_{\mbox{\tiny INT}}^2 = \frac 14$.

\section{Conclusion}
  As amply evident from the above discussion, the method just elaborated for constructing
vector coherent states is generic.
One could in this manner associate families of VCS to almost any hypergeometric function. More interestingly,
the method enables one to associate VCS to certain Clifford algebras and to reducible representations from the
principal series of locally compact groups. Some of these results will be presented in  forthcoming
publications.

\section*{Acknowledgements}
We would like to thank M. Bertola for interesting discussions. This work was partly supported by research grants
from the NSERC (Canada) and the FCAR (Qu\'ebec).


\begin{thebibliography}{XXXX}

\bibitem{key2} Ali, S.T., Antoine, J-P. and Gazeau, J-P., {\it Coherent States, Wavelets and their Generalizations,}
                   Springer-Verlag, New York, 2000.
\bibitem{AGMKP01} Antoine, J.-P., Gazeau, J.-P., Monceau, P., Klauder, J.R. and Penson, K.A., {\it Temporally
                  stable coherent states for infinite well and P\"oschl-Teller potentials\/}, J. Math. Phys.
                  {\bf 42}, 2349-2387 (2001)
\bibitem{key7} Bartlett, S.D., Rowe, D.J. and  Repka, J., {\it Vector coherent state representations,
                    induced representations and geometric quantization: I. Scalar
                    coherent state representation}, J.Phys. {\bf A35}, 5599-5623 (2002).
\bibitem{key8} Bartlett, S.D., Rowe, D.J., Repka, J., {\it Vector coherent state representations,
                     induced representations  and geometric quantization:
                      II. Vector coherent state representations}, J.Phys. {\bf A35}, 5625-5651 (2002).
\bibitem{bar-gir71} Barut, A.O. and Girardello, L., {\it New ``coherent'' states associated with non-compact groups\/},
                    Commun. Math. Phys. {\bf 21}, 41-55 (1971).
\bibitem{key9}  Borzov, V.V and  Damaskinsky, E.V., {\it Generalized coherent states for classical
                  orthogonal polynomials}, preprint math.QA/0209181 v1.
\bibitem{borzov01} Borzov, V.V., {\it Orthogonal polynomials and generalized oscillator algebras\/}, Integral
                   Transforms and Special Functions {\bf 12}, 115-138 (2001).
\bibitem{borzov95} Borzov, V.V., Damaskinsky, E.V. and Yegorov, S.B., {\it Some remarks on the representations of the
                   generalized deformed oscillator algebra\/}, preprint q-alg/9509022 v1; Zap. Nauch. Seminarov.
                   LOMI {\bf 245}, 80-106 (1997) (in Russian).
\bibitem{fujii01} Fujii, K., {\it Basic properties of coherent and generalized coherent operators revisited\/},
                   Mod. Phys. Lett. {\bf A16},  1277-1286 (2001).
\bibitem{key16} Gilmore, R., {\it Lie Groups, Lie Algebras, and Some of their applications}, John Wiley \&
                   Sons, New York (1974).
\bibitem{key19}Klauder, J.R and Skagerstam, B.S., {\it Coherent States, Applications in Physics and
                   Mathematical Physics}, World Scientific, Singapore, (1985).
\bibitem{odz01} Odzijewicz, A., Horowski, M. and Tereszkiewicz, A., {\it Integrable multi-boson systems and
                orthogonal polynomials\/}, J. Phys. {\bf A34}, 4353-4376 (2001).
\bibitem{odz98} Odzijewicz, A., {\it Quantum algebras and $q$-special functions related to coherent states
                maps of the  disc\/}, Commun. Math. Phys. {\bf 192}, 183-215 (1998).
\bibitem{key30}Perelemov, A.M., {\it Generalized coherent states and their applications},
                   Springer-Verlag, Berlin (1986).
\bibitem{key33}Rowe, D.J. and Repka, J., {\it Vector coherent state theory as a theory of induced representations},
                  J.Math.Phys. {\bf 32}, 2614-2634 (1991).

\end{thebibliography}
\end{document}